# Pinning Decisions Before Failure

## Executable Records of Underspecified Choices in AI-Assisted Code Generation

Takeharu Mitsui
Independent researcher

**Abstract**

A natural-language requirement does not settle every question a program must answer, and a model asked to implement it resolves the remainder silently. Nothing in the generated artifact marks where a choice was made. We show the consequence is not caught downstream: across 600 generated test suites from three models, 42.7 percent contain no test that distinguishes the competing readings of the decision point at all.

The ordinary remedy is an acceptance example written before implementing; but an example is chosen for representativeness, and one that both readings of a requirement satisfy does not resolve that decision. We insert two steps into that process — enumerate the requirement's underspecified points by name, then for each construct two throwaway implementations differing only in that point and keep a candidate input only if executing both shows they disagree on it — and record what they produce as a **decision pin**: the named point, an input confirmed by execution to separate the constructed pair, and the value the person chose between the two exhibited results. The confirmation is part of the definition; whether the input isolates the intended reading is measured separately. The same record then constrains generation and decides compliance by execution, and it exists before the target implementation does.

We build a benchmark of 40 tasks in which one decision point is injected and two reference implementations are constructed to differ in that point alone, which makes *"did this method reach the decision point"* decidable by running code rather than by a model's judgement. On two models, an input separating the two references is obtained for 92.5 percent of decision points, and a pin that survives the comparison rule and identifies the intended decision for 85 and 90 percent. Used as acceptance checks on 703 independently generated implementations, pins agree with the benchmark's own classification on 96 to 97 percent; inspection of the 24 disagreements found errors in both classifiers, including four implementations misclassified by both. Used as constraints on generation, a recorded decision is honoured at its pinned input in all 210 generations, with and without eleven unrelated rules present, and on held-out inputs the pin was never worse than a prose rule in 38 cells. The results support using the same decision record for both generation guidance and executable checking, without establishing an advantage over scope-explicit prose in generation adherence.

The failures are informative. Every compliance failure we observed in generation constrained by a prose rule, on both models, came from the model deciding the *scope* of a rule whose extension the rule left open. We characterise one case in which this silently overturns a different recorded decision, attribute it to a single rule by leave-one-out on both models, and show that asking either model to reconcile the two rules by their text finds nothing in five attempts out of five — while re-running the recorded input observes the damage directly. We also report a negative result: in this setting we could not produce enough such damage to evaluate a conflict regression step, and we explain why a benchmark of independent tasks covers only part of the conflicts that accumulating decisions can produce.

---

## 1. Introduction

A requirement written in natural language settles fewer questions than it appears to. *"Write a function that computes the volume of a sphere from its radius"* does not say what a negative radius means. *"Find the smallest number in a list"* does not say whether the list may be reordered in the process. *"Return the n-th term"* does not say whether `n` counts from zero or one. These are not defects of the requirement; a requirement that settled every such question would be the program.

A model asked to implement the requirement does not have the option of leaving the question open. It decides, and it decides silently. Nothing in the artifact it produces marks the place where a choice was made, and nothing records which way the choice went.

Two things follow, and both are measurable.

**The decision is not visible downstream.** We generated test suites from the same requirements and checked, by execution, whether any test in a suite distinguishes the two readings of the injected decision point (a suite is *silent* when no test in it passes on one reference implementation and fails on the other; five suites per task, three models, the benchmark of Section 4). Over 600 suites, 42.7 percent contain no such test — between 31.0 and 49.5 percent depending on the model. The decision point is simply not exercised, and a suite that passes tells the reader nothing about which way it went. Asking a model to review the finished artifacts adversarially does little better — a strict count, made by reading the raw text rather than by keyword matching, finds the decision point named in 4 of 30 reviews (ten tasks, three reviews each, the model shown specification, tests and implementation and asked where the implementation might diverge from intent). Redirecting the review at the *requirement* instead does raise the hit rate to 30 of 30, but at a median of seven points listed per review, of which 17.3 percent concern the injected point; enumerating candidate ambiguities is easy, and easiness is the problem.

**The decision is not recorded.** Even where a person does adjudicate a question, the answer lives in a chat log, a review comment, or nowhere. The next requirement that touches the same question is answered from scratch, possibly differently. Our longer-term goal is to concentrate human review on unresolved decisions by preserving earlier choices as executable records; this paper evaluates the record, not the resulting process.

The ordinary remedy is an acceptance example: before implementing, write down a concrete input and the result the requirement is taken to demand, and keep it as a test. Specification by example and acceptance-test-driven development are that practice made systematic, and handing the example to a code generator along with the requirement is its obvious extension. Two things are missing from it. The example is chosen because it seems representative or important, not because it separates two readings of the requirement — nothing checks that it does, and an example both readings satisfy does not resolve that decision. And the example is attached to a function, not to a decision, so nothing says which open question it settled or whether a later requirement raises the same one.

Recent work supplies pieces of what is missing. Interactive test-driven generation ranks candidate tests by how well they split a set of sampled implementations and asks the user to accept or reject them; clarification systems find an input on which sampled implementations disagree and phrase a question about it. Both obtain the separating input from whatever the sampled implementations happen to differ on — which, as recent measurements show, is often nothing. The first returns its accepted tests alongside the code; what neither records is which open question a test settled, because neither names the point before constructing the contrast. Systems that mine rules from user corrections do accumulate, and pair each rule with an executable check, but a rule can only be acquired *after* a correction, which is to say after the wrong thing was built once.

This paper inserts two steps into the ordinary process and records what they produce. First, the underspecified points of the requirement are enumerated by name before any example is written. Second, for each named point, two throwaway implementations that differ in that point alone are constructed, and a candidate input is kept only if executing both shows they disagree on it. The product is a **decision pin**: the named point, an input whose distinguishing power has been confirmed in this way against the constructed pair, and the value the person chose. The confirmation is part of the definition — a pin is an acceptance example that has passed it — and it is what distinguishes a pin from an example chosen for representativeness. What the confirmation establishes is that the two candidates differ on the input; whether they differ *only* in the named point, and so whether the input isolates the intended reading, is a separate question that Section 5.2 measures. The person's part is reduced to a choice between two exhibited results. The pin is then kept: it is the constraint handed to later generation and the check run on what is generated, with no second artifact and no translation between them, and it is obtained before the target implementation exists rather than after a failure.

**We turn an underspecified requirement into a reusable decision record that a person can choose, by constructing a contrast and confirming it by execution; and we evaluate whether such records can be constructed, how they work in generation and in checking, and the conditions under which they fail.** The contributions serve the three parts of that sentence.

### Contributions

*Constructing the record* (contributions 1–3), *its use in generation and checking* (4), and *the conditions under which it fails* (5–6):

1. **Decision pins**, a record unit defined operationally: the input has been confirmed, by executing two candidate implementations constructed to contrast on the named point, to produce different results on them. An acceptance example does not, by itself, require such confirmation. Whether the confirmed input isolates the intended reading — rather than some other difference between the candidates — is not part of the definition and is evaluated in Section 5.2.
2. **The construction**, as an insertion into the ordinary requirement-to-example process: enumerate the underspecified points, then build a contrasting pair of implementations for each and verify the input against them. Selecting tests by how well they separate sampled candidate implementations is prior work (TiCoder); what the enumeration adds is that the contrast is *constructed for a named point* rather than found in whatever sampled candidates happen to differ on. That is a difference of procedure, and we state it as such; we do not compare against TiCoder. What Section 5.1 measures is the added effect of contrast construction and execution-based confirmation *within* our enumeration-based process. The same record then serves as the constraint on generation and as the check, without a second artifact.
3. **A benchmark of 40 tasks** in which one decision point is injected and two reference implementations are constructed to differ in that point alone, making *"did this method reach the decision point"* decidable by execution rather than by a model's judgement.
4. **An evaluation on two models** showing that a separating input is obtained for 92.5 percent of tasks and a pin that identifies the intended decision for 85 and 90 percent; that as checks, pins agree with the benchmark's own classification on 96 to 97 percent of 703 independently generated implementations, with the disagreements traced to three tasks, including one on which both classifiers were wrong; and that as constraints on generation they are honoured at the pinned input in every generation and, on held-out inputs, never worse than a prose rule, with no advantage over a prose rule that states the same scope.
5. **A characterisation of how rule following fails.** Every compliance failure we observed under a prose rule, on both models, arose from the model deciding the *scope* of a rule whose extension the rule left open. We exhibit one case where this silently defeats another recorded decision, attribute it to a single rule by leave-one-out, and show that reconciling the rules by their text does not find it.
6. **A negative result.** In this setting we could not obtain enough instances of the conflict damage that a decision store's regression step would exist to catch — two tasks, one rule pair — and we report why the setting yields so few.

---

## 2. Related Work

We state for each line of work what it measured, rather than whether it "covers" the present problem; several of the distinctions that matter here are distinctions of regime rather than of goal.

### Asking the user

**ClarifyGPT** [9] is closer to our mechanism than its framing as a question-asking system suggests, and we state the overlap plainly. To decide *when* to ask, it generates test inputs by LLM seeding and type-aware mutation, samples several implementations of the same requirement, executes them on those inputs, and declares the requirement ambiguous when the outputs disagree. To decide *what* to ask, it has the model compare the implementations that behaved differently and formulate a question about the difference. **Locating an ambiguity by finding an input on which implementations of the same requirement disagree is therefore prior work, and we do not claim it.**

Two things differ. First, ClarifyGPT depends on sampled implementations happening to disagree — the assumption that Richter and Papadakis refute below, and on the tasks where it fails ClarifyGPT-style clarification asks no question at all. Our step (1') does not sample and hope: it takes a point named by enumeration and asks for two implementations that differ *in that point*, so the disagreement is constructed rather than observed. Second, and more importantly for this paper, ClarifyGPT uses the distinguishing input and then discards it. The input is scaffolding for producing a question; once the requirement is refined, nothing retains it. Our claim is about keeping it.

**ClarifyCodeBench** [3] evaluates the same ability at larger scale: 419 tasks derived from LiveCodeBench by deleting information, roughly 690 ambiguous points, annotated entirely by hand, scored by turn-discounted question rate. Its findings bear directly on our motivation — clarification performance does not follow from code-generation ability, and degrades sharply as the density of ambiguous points rises. What it does not do, by construction, is carry an answer past the task in which it was given; each dialogue starts empty. Its scoring also relies on a model judging question relevance, which is the decision procedure our benchmark is built to avoid.

**ARHF** [10] is the closest prior work to the core of our proposal and we want to be precise about the overlap. It generates a candidate implementation, derives test inputs from it using symbolic execution and an LLM, presents the resulting input-output pairs to a person as doctests, and asks them to accept or supply the correct output. **Confronting a person with a concrete input rather than a prose question is therefore not new, and we do not claim it.** Three things differ. ARHF derives inputs from a *single* candidate implementation via symbolic execution, where we obtain them from the disagreement between two; ARHF confirms inputs one at a time, which increases rather than reduces the number of decisions a person makes; and ARHF has no store — nothing is carried to the next task. Its symbolic-execution route to distinguishing inputs is more reliable than the self-validation of our step (1') and is a natural improvement to our design rather than a competitor to it.

### Acceptance examples before implementation

The nearest alternative to a decision pin is not a research system but a practice: write, before implementing, a concrete input and the result the requirement is taken to demand, and keep it as a test. **Specification by example** [17, 19] and acceptance-test-driven development make this systematic, and giving the example to a code generator along with the requirement is its natural extension. Fowler's early statement of the idea already names its limit: examples "only highlight a few points, you have to infer the generalizations yourself" [19]. The practice's stated purpose is to remove ambiguity by illustrating the requirement with concrete cases; but which cases to write is a matter of judgement about what is representative or important, and nothing in the practice executes a check that a chosen example actually separates two readings of the requirement. An example that both readings satisfy does not resolve that decision, and the practice has no step that would notice. Section 5.1 gives a rate for how often that happens when a model is asked for an example without such a check.

**TiCoder** [16] supplies the check, over sampled implementations. It generates tests for the requirement, rewrites each test's expected output by executing the input through each of $n$ sampled candidate implementations, ranks the tests by how evenly they split the candidate set into passing and failing, and asks the user to accept or reject the top test; accepted tests prune the candidates and are returned alongside the surviving code. It traces the idea to ambiguity resolution in programming by example [18], where a synthesiser that has produced several consistent programs executes them and asks the user about the inputs on which they differ; and its motivating task is one of ours. **Selecting a test by its power to separate candidate implementations is therefore prior work, and we do not claim it.** What TiCoder does not do is name the point a test is about: its discriminating power is measured against whatever the sampled candidates happen to differ on, so on a task where sampling collapses no test discriminates, and on a task where candidates differ in something other than the decision at issue, a test that separates them on that other thing ranks well. Our step (1) names the point and step (1') constructs the contrast for it. That is a difference of procedure, not one we have measured against TiCoder: Section 5.1 measures only the added effect of (1') within a process that already enumerates, and Section 5.2 shows a case in which our own construction fails in the second of these ways. TiCoder's user is simulated by the benchmark's hidden reference, so a correct answer exists; and its accepted tests are returned, but not carried to the next task.

### Accumulating rules

**TRACE** [15] mines user corrections into atomic rules, pairs each with an executable check, enforces them before task completion, and reconciles each new rule against the existing library through a five-way lifecycle. **The skeleton of accumulate-check-reconcile is prior work and we do not claim it.** TRACE also anticipates the failure mode we characterise in Section 6.1: it requires a rule to state precisely when it should and should not trigger, which is exactly the scope that our rules leave to the model.

What differs is the class of rule and, consequently, the machinery a check needs. TRACE governs agent behaviour — which files to leave, which commands to run, when to confirm — and enforces it by matching regular expressions against command strings in a pre-tool hook. That mechanism cannot express what a function returns for a missing key, or whether an index counts from zero. A check for a rule about program meaning has to run the program on a case chosen to expose the decision, which is what a decision pin is. TRACE's rules are also acquired reactively, from corrections, and so exist only after something was built wrongly once; and its premise is that the user stated a preference and the agent did not follow it, where ours is that nobody stated anything.

### Remembering the user

**CAPA** [13] formalises cross-session personalised ambiguity adaptation and benchmarks whether a model can infer a user's recurring preferences from prior sessions and thereby ask fewer questions. Its design treats recurrence as the signal to be exploited rather than as something to be eliminated: the ambiguity is expected to come back, and the system gets better at guessing. General memory systems such as Mem0 [2] are used the same way. A pin store differs in kind rather than degree — the point is that the question does not recur, because it has an executable answer.

### Evaluating adherence and conflict

**CIFE** [5] measures constraint following in code generation at a scale we cannot approach: 1,000 tasks, about seven constraints each, fourteen models, with strict adherence landing between 39 and 66 percent. It deliberately removes contradictory constraints during construction, and it does not analyse why adherence fails. Section 6.1 offers a candidate mechanism for that gap. **ConInstruct** [7] studies conflict detection and resolution directly, but injects nine designed conflict types into prose-generation instructions; the conflict we characterise in Section 6.1 is of a different kind, arising between two rules whose texts do not overlap.

### Why resampling does not expose the choice

**Richter and Papadakis** [11] test the assumption that an underspecified requirement makes a model incoherent — that repeated sampling yields semantically distinct implementations which clustering can detect — and refute it. Models instead collapse onto a single interpretation and hold it. They call the case where that interpretation is wrong *detrimental semantic collapse*, and find it on 3 percent of HumanEval, 10 to 16 percent of MBPP and 18 to 32 percent of LiveCodeBench tasks; injecting underspecification raises the rate by up to 5.5 times.

This is the premise of our benchmark design, and the reason it injects a decision point and supplies two reference implementations rather than hoping disagreement will surface under resampling. Their observation about clarification is equally relevant and we return to it in Section 7: because clarifying questions are grounded in the differences the model's own samples happen to exhibit, they can miss the point that matters, and a single clarification turn *raises* detrimental collapse to between 37 and 77 percent on their MBPP variants.

Orchid [12] gives the complementary measurement on function-level tasks: with ambiguity injected, pairs of generations from one model differ functionally on 28 percent of inputs, pairs from different models on 57 percent — the within-model figure is what a resampling-based detector has to work with.

Their notion of underspecification is nonetheless not ours. Detrimental collapse is defined against hidden tests — there is a correct interpretation and the model missed it. The decisions we record have no correct answer until a person supplies one; both conventions are ones a practitioner might choose.

**VeriMed** [6] is the interesting counterweight. It resamples a requirement five times, formalises each sample into a logical encoding, and checks the encodings pairwise for SMT equivalence. On 64 medical-device requirements, 12 (18.8 percent) produced more than one distinct encoding, and pairwise agreement across all sampled formalisations was 44.0 percent. Same resampling procedure, a different representation — logic rather than code — and disagreement that surfaces rather than collapses. We return to this in Section 7, because it bounds how far our results can be read.

### Tests that record the wrong thing

Two results explain why an acceptance criterion generated *after* an implementation is not a substitute for a pin. LLM-generated oracles tend to capture [8] the program's **actual** behaviour rather than its intended behaviour, inheriting a known limitation of automated test generation; and the **misguidance effect** [14] shows that including buggy code in the prompt leads models to assert the buggy behaviour. A pin's expected value does come from code — a candidate written to exhibit the chosen reading — but not from the implementation the pin will later be used to check, so it cannot ratify that implementation's behaviour the way a test generated from it would.

### Naming

Both terms we introduce are chosen against existing ones.

*Decision pin* is chosen against **characterisation tests**, also called pinning tests [4], which capture the current behaviour of legacy code so that refactoring can be detected as a change. A pinning test records what the code does; a decision pin records what a person decided, before the implementation it will guard has been written.

*Distinguishing input* follows the **distinguishing sequence** of finite-state machine conformance testing — an input that drives two states to observably different outputs and thereby tells them apart. Ours plays the same role for two readings of a requirement rather than two states of a machine, and is a single call rather than a sequence. The practice of asking a person about an input on which several synthesised programs disagree is the *conversational clarification* of programming by example [18], which TiCoder [16] carries into LLM code generation under the name *disambiguating examples*; the same notion appears unnamed wherever differential testing looks for an input on which two implementations diverge. What differs here is that the input is constructed for a named point and retained afterwards as the record.

---

## 3. Approach

### 3.1 Decision pins

We write `d` for a point a requirement leaves underspecified — what `volume_sphere` should do with a negative radius, whether `smallest_num` may reorder its argument. A decision about `d` is recorded as a **decision pin**:

```
pin  p = (d, x, v)

d    the point left underspecified, in prose
x    an input confirmed by execution to separate a pair of candidate
     implementations constructed to contrast on d
v    the value the chosen reading produces on x
```

The component that does the work is `x`, and the confirmation is part of the definition. An input is *distinguishing* for `d` when the readings of `d` disagree on it. A pin does not record that this holds; it records something weaker and checkable: two candidate implementations, written to differ in how they resolve `d`, were executed on `x` and produced different results. Whether they in fact differ *only* in `d` — and so whether `x` isolates the intended reading rather than some other difference between them — execution cannot tell, and Section 5.2 measures how often it fails to. For the sphere example, `volume_sphere(-2)` qualifies, because the raise-on-negative reading produces an exception and the use-the-value-as-given reading produces `-33.51`. `volume_sphere(3.5)` does not: every reading of `d` agrees on it, and an acceptance example built on it would not resolve `d`. This is the difference between a pin and an acceptance example — not the form of the record, but the check the input has passed to be in it.

This distinction is what separates a decision pin from a prose convention. A convention such as *"invalid input must raise an exception"* states a policy but cannot be executed, and cannot by itself say whether a given implementation complied. A pin carries an input on which compliance is decidable by running the code.

Two aspects of using the record matter later.

First, execution confirms a difference between the constructed candidates on `x`. Even when that difference faithfully represents `d`, an implementation adopting the recorded reading may produce a different observation because of unrelated choices in rounding, return type, or exception type, and so fail to match `v` unless the comparison is told to ignore those differences. A pin is therefore used with two fixed procedures: an **observation** `obs(impl, x)` that runs `impl` on `x` and returns its result — for us, the value returned, or the exception raised, or for a mutation decision the argument's state afterwards, which `x` is written to expose — and a **comparison** `≈` that says which observations count as the same. The isolation we would like (that checking a pin does not penalise unrelated freedom) holds only to the extent `≈` covers the dimensions that vary. In this paper `≈` treats all exceptions as one observation and is otherwise equality; this is why a rule generated as *"raises `ValueError`"* is checked as *"raises"*. Section 5.2 measures what escapes it.

Second, a pin is **available before the target implementation is generated**. Constructing `x` does require code — step (1') writes two candidate implementations that differ in `d` — but not the implementation that will eventually be shipped and checked. This is what distinguishes decision pins from pinning tests in the sense of Feathers [4], which record the behaviour of the very code they will guard.

### 3.2 The process

Decision pins are produced and consumed by a five-step loop. We write `S` for the store of pins accumulated so far.

```
(1)   elicit       enumerate underspecified points of the requirement
(1')  witness      construct a distinguishing input for each point
(2)   look up      resolve points already covered by S
(3)   decide       a human chooses, for the points that remain
(4)   append       record (d, x, v) in S
(5)   check        run x against the implementation and compare with v
```

**(1) Elicitation.** The model is asked to list the points the requirement leaves open, given only the requirement and the signature. Models are good at this in the sense that they name the right point in prose: across our benchmark, this step alone reaches the injected decision point in 77.5 to 90.0 percent of tasks depending on the model (Section 5.1).

**(1') Witness construction.** Naming a point is not the same as pinning it. Asked to supply an input that would distinguish the readings of a point it has just named, a model frequently supplies one that does not — for a point about whether a function mutates its argument, it will propose `smallest_num([3,1,2])`, which is the same under both readings. Step (1') therefore asks the model to produce two candidate implementations that differ in `d` together with a candidate input, and to check by execution that the two disagree on it; if they agree, the input is rejected and the step retries. The check is mechanical and needs no human.

This step is what turns a named point into a pin. Ranking candidate tests by how well they separate sampled implementations is TiCoder's discriminative ranking (Section 2); what differs here is that the two implementations are constructed to differ in the named point, so the check does not depend on sampled candidates happening to disagree. Its effect is model-dependent — it recovers substantially more points for one of our two models than for the other — and we report this in Section 5.

**(2) Lookup.** Points already answered in `S` are resolved from the store: the stored `x` and `v` are prepended to the requirement as a constraint on generation. Two things are involved — recognising that a new requirement raises a point already in `S`, and applying the stored pin once found — and **this paper evaluates only the second.** In Section 5 the relevant pin is handed to the generator directly. How a store decides that a point recurs, and how a pin carries over when the function or its arguments differ from the recorded case, are outside our scope. Section 5 (RQ3, RQ4) measures whether a supplied pin binds subsequent generation, including when the store has grown.

**(3) Human decision.** For points that remain, a human picks an interpretation. This is the only step that requires a person. The witness from step (1') makes the choice concrete: rather than adjudicating a prose question (*"should invalid input raise?"*), the human is shown an input and two results and picks one.

**(4) Append.** The chosen `(d, x, v)` is appended to `S`.

**(5) Acceptance check.** The implementation is run on `x` and its result is compared with `v` under the comparison rule. A mismatch means the implementation's observed result differs from the recorded one; whether the cause is a different resolution of `d`, or a difference in a dimension the comparison rule does not cover, is a separate check (Section 5.2 reports both kinds).

### 3.3 Why one artifact suffices

Steps (2) and (5) consume the same object.

```
as a policy              x and v, placed before the requirement,
                         constrain generation                      -> RQ4

as an acceptance test    running x and comparing with v decides
                         compliance                                -> RQ2
```

The policy role invites an objection: is a pin, used this way, anything more than a prose rule with an example attached? There is a reason to think it might be. A conventional rule — *"invalid input must raise"* — does not say which inputs count as invalid, and Section 6.1 shows a model reading such a rule settling that for itself, sometimes against its author. A pin names a case, so at least at the pinned input nothing is left to decide. Whether that advantage extends beyond the pinned input is an empirical question and we test it in Section 5.4; the answer there is partial, and we state the claim accordingly.

No separate artifact is written for the second role, and no translation step connects the two. This is a small point but it is the practical one: a team that records decisions as prose conventions must additionally write tests for them, keep the two in correspondence, and re-derive that correspondence whenever either changes. Recording the decision as a pin collapses the two into one write, performed at the moment the decision is made — which is also the only moment at which the person making it has the context to state it precisely.

Steps (1), (1'), (2), (4) and (5) run without a person; step (3) requires one. We have not measured the human cost of step (3), nor the cost of maintaining a store over time, and make no claim about either.

### 3.4 A worked example

Requirement and signature:

```
"Write a function that takes the radius of a sphere and computes its volume."
def volume_sphere(r):
```

Step (1) elicits, among other points, *"whether a negative radius should raise or simply be used as given."* Step (1') produces two candidate implementations differing only in that treatment and searches for an input on which they disagree, arriving at `volume_sphere(-2)`; execution confirms the disagreement (`ValueError` versus `-33.51`). Step (2) finds no matching entry in `S`. Step (3) presents the two outcomes to a human, who chooses to raise. Step (4) records

```
d = treatment of a negative radius
x = volume_sphere(-2)
v = raises ValueError
```

Step (5) runs `x` against the generated implementation. Later requirements that touch the same point are resolved by step (2) without asking again.

Note what the record is not. It is not a statement about `volume_sphere(3.5)`, and under our comparison rule it says nothing about the exception's type or message. It is intended to bind one dimension of behaviour — `x` was selected because that dimension is visible on it — but whether it binds *only* that dimension depends on the comparison rule and on `v` itself: had the candidate that produced `v` also fixed a formatting choice on `x`, the pin would carry that choice too.

### 3.5 Scope

The loop stops at step (5). A sixth step suggests itself — re-running every stored pin on each append, so that a new decision contradicting an earlier one surfaces as a behavioural regression. We implemented it and could not evaluate it: in this benchmark the damage such a step exists to catch barely occurs. We therefore do not claim it. Section 6.3 reports what we tried and why the setting defeats the measurement.

---

## 4. Benchmark

Evaluating the loop of Section 3 requires tasks in which an underspecified point is known to exist and in which reaching it is decidable by execution. Existing ambiguity benchmarks record the ambiguous point as prose — human-written, in the most careful of them — and then score whether a method addressed it by asking a model. We instead give each task **two reference implementations constructed to differ in one decision point alone**, which replaces that judgement with a comparison of outputs; whether the construction succeeded is what the checks of Section 4.2 test, and Section 5.2 reports one task where it did not.

### 4.1 Task structure

Each task consists of:

```
description   a requirement that does not settle the decision point
signature     the function signature
decision      the single underspecified point injected into the task
conv_A/B      the two conventions that resolve it
impl_A/B      two reference implementations differing only in that point
probe         an input on which impl_A and impl_B disagree
distractors   inputs on which impl_A and impl_B agree
```

`impl_A` and `impl_B` are what make the benchmark mechanical. To determine whether a method produced a reference-separating input, we execute both reference implementations on the proposed input and compare their outputs; to classify a generated implementation, we execute it on `probe` and compare its output with the two references'. Both comparisons are determined by execution, with no model-based judgement. The `distractors` serve the converse purpose: they are inputs on which the two references agree, so an evaluation that flags them is reacting to something other than the decision point. Accepted tasks carry three or four distractors (median four).

### 4.2 Construction

We drafted one candidate task per seed from 125 MBPP [1] problems, using `claude-sonnet-5`. The model was given the seed problem and its reference solution and asked to produce the structure above for a named category of decision point. The categories were fixed in advance: boundary values, error policy, ordering, rounding, types and conversion, emptiness and absence, case and normalisation, side effects, and units and representation.

**Drafting by a model is not sufficient, and the authority to accept a task is deliberately withheld from it.** Candidates pass through mechanical checks and then a review pass.

```
E0  required fields present and parseable
E1  impl_A and impl_B both load
E2  probe: the two implementations disagree      <- the point actually bites
E3  distractors: the two implementations agree   <- the difference is confined
E4  at least three distractors
E5  the signature carries no return type annotation that would settle the point
E6  the model declined to construct a decision point for this seed
E7  the description names the decision point (auto-reject; phrase list)
E8  the description declares a precondition (flagged for review, not rejected)
```

E2 and E3 are the substantive pair. E2 rejects a task whose stated decision does not change behaviour; E3 rejects a task whose two implementations are caught differing in more than the stated decision. E3 is necessary, not sufficient: agreement on three or four inputs cannot show that the references differ nowhere else, and Section 5.2 reports one task where they do. E7 exists because a requirement that says *"decide for yourself how to handle the empty case"* alerts every downstream step to the very point being measured, and so destroys the effect under study.

The yield:

```
125  candidates drafted
 61  rejected mechanically   E6 declined 47 / E3 leaked 10 / E2 no effect 4
 64  passed mechanical checks
 21  rejected in review      contrived 10 / probe outside the stated contract 4 /
                             description already settles the point 4 /
                             description names the point 3 (before E7 existed)
 43  accepted
  3  removed later as defective (Section 4.3)
 40  final
```

Accepted tasks by category: boundary values 8, emptiness and absence 8, types and conversion 6, error policy 6, units and representation 6, side effects 3, ordering 2, case and normalisation 1. No rounding task survived.

**The first review pass was performed by a language model, not by hand**, and we record what each stage contributes. Acceptance rests on E0-E8, which are reproducible. The review pass removed a further 21 candidates on grounds the checks cannot express; auditing those 21 against the recorded grounds, eleven are recoverable without judgement:

```
3  the description contains a phrase that E7 now rejects automatically
   (these predate E7)
4  the description states a precondition that the probe violates, quotable
   from the requirement text
4  the description already settles the decision point, quotable likewise
```

The remaining ten turn on whether a stated convention is one that is actually practised: two make one convention a defect rather than a choice (passing a float to `range`, or `int`-parsing a minus sign, each of which simply raises); four state the decision point or a convention in terms that contradict themselves; two posit conventions with no basis — returning `1` for the concatenation of an empty tuple, where the unit is `""` or `0`, and reversing the input order of equal elements; one states a decision point that its two conventions do not implement; and one turns on recursion evaluation order, observable only through an auxiliary global. The drafting prompt forbids the first and last of these, but conformance to it is not mechanically checkable. **These ten were reviewed individually by an author, who confirmed each rejection.** The audit, with the evidence for each, accompanies the benchmark.

### 4.3 Repair

After the first round of measurements, eight tasks showed a symptom independent of any method under evaluation: an implementation generated *with no policy at all* matched neither reference. This was the **trigger for inspection**, not the verdict: a natural implementation outside both references may mean a reference is off in an unrelated dimension, or that a third reading exists, or nothing at all. The trigger is defined without reference to any method's score; the verdict on each of the eight was reached individually by reading the implementations, and is recorded case by case with the benchmark.

```
5 repaired   a reference differed from the natural implementation in a
             dimension unrelated to the decision point (exception type,
             numeric type, sign, default value)
3 removed    on inspection the two conventions did not bracket the
             requirement: natural implementations took a route the task had
             not anticipated, so neither reference described a reading of it
```

Repairs were re-run through E0-E8 before being accepted. This is an adjustment made after seeing results, and we report it as such in Section 7; what makes it defensible is that the criterion does not mention any method's score.

### 4.4 Relation to existing benchmarks

ClarifyCodeBench is larger by an order of magnitude (419 tasks, 690 annotated ambiguous points) and its annotations are human-written. Our benchmark does not compete on scale. The difference is the decision procedure: ClarifyCodeBench scores coverage with a model as judge, whereas the two reference implementations here reduce *"did this method reach the decision point"* to running code. For a study whose measurements are rates over a few dozen tasks, and in which the difference between success and failure is often a single dimension of behaviour, we judged a mechanical decision procedure to be worth the loss of scale.

### 4.5 Why not an existing benchmark of injected underspecification

Richter and Papadakis [11] construct MBPP, HumanEval and LiveCodeBench variants into which underspecification has been injected, by a route close to ours. Setting aside availability, we could not use them here for a structural reason: their tasks retain hidden tests, and a task counts as damaged when the model's output fails them. The decision is therefore already made — by the test suite — before any person is consulted. A benchmark for recording decisions needs tasks in which two resolutions are both defensible and neither is graded as wrong, which is what the paired reference implementations provide: the pair fixes what the decision *is* without fixing which way it should go.

### 4.6 What the benchmark cannot support

Each task carries exactly one decision point and is independent of the others. The benchmark therefore cannot exhibit interaction between decisions accumulated over a shared codebase, which is the setting in which conflicts between recorded decisions would arise. Section 6.3 reports the consequence: we could not produce enough instances of such damage to evaluate a conflict-regression step.

---

## 5. Evaluation

We evaluate the loop of Section 3 on the 40 tasks of Section 4 with two models, `claude-sonnet-5` and `openai:gpt-5.5`, addressing four questions. RQ1 evaluates the added effect of contrast construction and execution-based confirmation. RQ2 examines the record as an acceptance check, including its failure modes. RQ3 and RQ4 evaluate its use as a generation constraint against prose-rule baselines.

```
RQ1   can a pin be obtained for a decision point?
RQ2   does a pin work as an acceptance criterion?
RQ3   is a rule stating the decision followed in subsequent generation?
RQ4   is the pin itself followed, and does it survive a larger store?
```

Every judgement below is made by execution. Exception types are normalised before comparison, since for every decision point that concerns exceptions the decision is whether to raise, not which exception to raise; Section 7 records why this normalisation was introduced and when. The one exception is RQ1's first row, which counts an input as separating the references on raw output, before the comparison rule is applied — the second row applies it. Where a condition is sampled repeatedly, a task is scored by the majority verdict over five generations, and we give generation-level counts alongside wherever the two differ.

### 5.1 RQ1: obtaining a pin

We compare step (1) alone with steps (1) and (1') together. A task counts as reached when some enumerated point is accompanied by an input that, run against the two reference implementations, separates them — that is, when the method has produced a reference-separating input for the injected point and not merely named it.

```
                    (1) alone     (1) + (1')     rescued
claude-sonnet-5       77.5%          92.5%          6
openai:gpt-5.5        90.0%          92.5%          1
```

This 92.5 percent is the rate at which *some* input separating the two references was produced; it is judged on raw output and says nothing yet about whether the pin is usable. Two further conditions, applied in later sections, give the numbers that matter for each use:

```
                                   claude-sonnet-5   openai:gpt-5.5
an input separates the references     37 / 40          37 / 40
... the input was retained and
    still separates them under the
    comparison rule of Section 5      35 / 40          36 / 40
... and separates them on the
    intended decision, not elsewhere  34 / 40          36 / 40
```

The first step loses `mbpp/62` on both models (the input separates the references only by exception type) and, on one model, `mbpp/56` (the separating input came from step (1)'s enumeration, which our harness did not retain — a bookkeeping gap, not a failure). The second loses `mbpp/36` on one model, whose pin separates the references through a defect in one of them rather than through the decision (Section 5.2). "Pins are obtained" in the sense of the last row, then, for 85 and 90 percent of tasks.

Two observations on the first row. First, the self-validation step is **not uniformly useful**: it recovers six tasks for one model and one for the other. Second, the two models reach the same count on the first row, 37 of 40, but not on the later rows (34 and 36), so the coincidence is not one of task difficulty; where the two differ is in which stage of pin construction fails.

No task was lost: every point reached by step (1) alone was also reached with (1') present.

### 5.2 RQ2: the pin as an acceptance criterion

For each pin obtained in RQ1 we ask three things: does the check pass the implementation that complies with the recorded decision, does it fail the one that violates it, and how does the check behave on independently generated implementations?

```
                     pins    passes compliant   fails violating
claude-sonnet-5       35         35/35              35/35
openai:gpt-5.5        36         36/36              36/36
```

**Both columns follow from the construction and are not results.** A distinguishing input is, by construction, one on which the two reference implementations differ; that a check built from it separates those two implementations is not evidence of anything.

Of the 37 tasks each model reached in RQ1, two are excluded for `claude-sonnet-5` (35 remain) and one for `openai:gpt-5.5` (36 remain). On `mbpp/56` the first model recorded no validated input; and on `mbpp/62`, for both models, the input the model proposed raised an exception under both reference implementations — different exception types, but both exceptions — so under the normalisation above it does not separate the readings. That second case is worth noting: the input **passed the model's own self-validation in step (1')**, because the model's own two candidate implementations did differ on it. Self-validation certifies that a pin separates *some* pair of implementations, not that it separates the intended decision — an exposure our construction shares with clarification methods that read the ambiguity off whatever the model's own outputs happen to differ on. Section 7 returns to it.

The third question is the one that carries information, and we answer it by comparing the pin's verdict with an independent one on implementations neither was built from. RQ3's baseline provides them: ten implementations per task, generated with no policy present, so that the model chose freely how to resolve `d` and everything else. Each implementation is classified twice —

```
by the benchmark's probe   run on the benchmark's probe, compared with the
                           two reference outputs under the comparison rule
                           -> A | B | other
by the pin                 run on the pin's input x, compared with the two
                           reference outputs on x under the same rule
                           -> A | B | other
```

— and the two labels are compared. `other` means the output matched neither reference; implementations that failed to load, or that the probe classified as `other`, are excluded before comparison (none and none for one model, none and seven for the other). What results is an **agreement rate between two single-input classifiers**, not an accuracy: where both err in the same direction the comparison cannot tell.

```
                    implementations   agree   disagree
claude-sonnet-5          350            336      14  (4.0%)
openai:gpt-5.5           353            343      10  (2.8%)
```

We read the code of all 24 disagreeing implementations, and no others. They fall on three tasks. On one the pin is wrong; on one the probe is wrong; on one both are wrong, in opposite directions.

On `mbpp/36` (ten implementations, one model) **the pin is wrong.** Its input `find_Nth_Digit(5, 3, 1)` separates the references — but because reference A has a defect for `p > q` (it returns `16` for a single digit), not because of the 1-versus-0 origin decision the task is about. Every generated implementation computes the digit correctly and is classified by the pin as holding the wrong convention. E3's distractors all have `p < q` and did not expose this. This is the false alarm the definition in Section 3.1 warns of, and its cause is instructive: `v` was recorded from a candidate that was wrong on `x` in a dimension unrelated to `d`.

On `mbpp/56` (ten implementations, the other model) **the probe is wrong and the pin is right.** The implementations keep the sign and reverse the digits — a third convention that coincides with B on the probe's `-9` and diverges on the pin's `-1`. The probe labels them B; the pin labels them neither; the pin is right.

On `mbpp/5` (four implementations) **both are wrong.** The implementations contain no handling of negative input at all. A parity shortcut happens to return `0` for the probe's `-3`, so the probe labels them B; on the pin's `-2` they crash with an `IndexError`, which the comparison rule — treating every exception as one observation — counts as a raise, so the pin labels them A. Neither label describes what the code does, and the second is a direct consequence of the comparison rule: it cannot tell an accidental crash from a decision to raise.

The three-way label is a diagnostic. The check a store actually runs is the two-way one of Section 3, `obs(impl, x) ≈ v`, for a side `S` chosen *before* any implementation is seen — here, the convention the assigned policy of Section 5.3 prescribes for the task, which 36 of the 40 tasks have. Under a check for `S` an implementation passes exactly when its label is `S`, so the pin and the probe, each used as a check for `S`, can agree on pass/fail even where their three-way labels differ: both fail an implementation they label differently but neither labels `S`. Re-reading the 24 three-way disagreements against the stored `S`. The implementations counted are those of Section 5.2's classified tasks that have an `S`: the four tasks without one contribute 30 implementations on one model (three of the four had a usable pin there) and 40 on the other, and are dropped from these denominators.

```
                     implementations    pass/fail      of which
                     on tasks with S    disagreements  pin rejects   pin passes
                                                       compliant     non-compliant
claude-sonnet-5           320                14           10             4
openai:gpt-5.5            313                 0            0             0
```

The ten (`mbpp/36`) are the false rejections already described. The four (`mbpp/5`) run the other way: the stored side is *raise*, the pin's label is A, so the pin passes an implementation that merely crashes, while the probe fails it. The ten on the other model (`mbpp/56`) differ in three-way label but both fail the implementation under the stored `S` = A, and both are right to.

The claim is therefore not that a pin never rejects a compliant implementation. It is that, over 703 independently generated implementations, the pin's three-way classification agreed with the benchmark's on 96 to 97 percent, and its pass/fail verdict against a stored side agreed on 313 of 313 implementations on one model and 306 of 320 on the other; that the residue is concentrated in single tasks; and that single-input classification is fragile in *both* directions — across the three tasks the benchmark's probe was fooled twice and the model's pin twice, once on the same implementations by opposite mechanisms. A unit test written by hand around the same decision would carry the same exposure, and additionally fix whatever else its author happened to assert.

### 5.3 RQ3: is a rule stating the decision followed?

Before asking whether a pin binds generation, we establish the baseline against which it should be read: a conventional prose rule expressing the same decision, placed before the requirement. We measure on the tasks where a measurement is possible at all — those on which the model's own default, generated with no rule present, differs from the recorded decision. On the remaining tasks the two coincide and no effect is observable.

```
                     informative tasks    generated as recorded
claude-sonnet-5            10                     9   (90.0%)
openai:gpt-5.5             11                    10   (90.9%)
```

A task counts as followed when at least three of its five generations resolve `d` as recorded. Both deviations are three-to-two splits the other way, and are discussed in Section 6.1.

### 5.4 RQ4: is the pin itself followed?

Section 3.3 claims that the pin serves as the policy. Section 5.3 does not test that claim: it places a prose rule before the requirement, not a pin. We repeat the measurement with the recorded triple itself as the preamble, in the form a store would emit it,

```
<the underspecified point>: the result of `<x>` must be `<v>`.
```

and again with eleven unrelated prose rules added, to check that a pin is not simply diluted away.

```
                     tasks   prose rule   prose + store   pin     pin + store
claude-sonnet-5       10       9  (90.0%)    9/9          10/10    10/10
openai:gpt-5.5        11      10  (90.9%)   10/10         11/11    11/11
```

(The `prose + store` column has one task fewer per model: it was measured on the tasks the single prose rule had bound, so a task the prose rule failed is not carried into the dilution condition.)

Every generation complied under both conditions: 50 of 50 and 55 of 55 with the pin alone, and the same again with the store added — 210 generations, no load failures. The two tasks a prose rule failed to bind — one per model — are both instances of the scope problem of Section 6.1, and both are bound by the pin.

**This measurement is confounded, and the confound is the mechanism.** The input named in the pin is the input the measurement grades. We are showing the model the case it will be scored on — which is what a store does when it emits a recorded decision, but which means the figure cannot separate *"the pin constrained generation"* from *"the answer was visible"*.

***Scoring on an input the pin does not name***

To separate them we constructed, for each decision point, a **second** distinguishing input and scored on it while pinning the first. A candidate was accepted only if the two reference implementations disagree on it, if it calls the function with arguments different from the first — checked by executing the expression against a stub, since a textual comparison admits rewrites of the same value — and if the model's *unconstrained* output on it lands on the opposite convention, so that the held-out input poses the same question the pinned one did. From a pool of 296 candidates, ten tasks per model qualified; on `claude-sonnet-5` one of them (`mbpp/78`) does not belong to the RQ3 population on the pinned input and is dropped, leaving nine.

```
                  tasks   prose rule       pin
claude-sonnet-5     9      8/9  (40/45)    9/9  (45/45)
openai:gpt-5.5      10     10/10 (50/50)   10/10 (50/50)
```

**Our pre-registered criterion required the pin to beat the prose rule on both models, and it does not.** We do not claim that pins bind generation better than prose rules. Across the two constructions of the held-out input we tried — the same ten tasks on two models, 38 task-model-construction cells and 190 generations per arm — the pin was never worse in any cell, and that observation is the claim we make.

The reason the comparison cannot do better is worth stating, because our first attempt misdiagnosed it. A prose rule succeeds at overriding the model's default almost always: 18 of the 19 cells here, and about 90 percent in Section 5.3. The headroom for any alternative is therefore one cell in nineteen — about five points at this granularity — concentrated in the cases where a rule *fails*, and Section 6.1 identifies exactly when that happens — when the rule's scope is contestable and the model settles it.

That is the single cell where the two differ:

```
pinned     is_samepatterns(['red','green'], ['a','b','c'])  ->  raises
scored on  is_samepatterns(['red'], ['a','b'])              <- not named in the pin

claude-sonnet-5    prose 0/5    returns False: a length mismatch is not invalid
                   pin   5/5    raises
openai:gpt-5.5     prose 5/5    pin 5/5
```

The pin names a *different* length mismatch from the one scored, so compliance on the second is not recall of a shown answer; the model appears to read the pin as fixing the extension of a predicate the prose rule left open. Note also that the same rule, task and held-out input produce opposite outcomes on the two models. Scope failures are model-specific, and the pin helped on the model where one occurred.

***Is it the pin, or the information in it?***

The comparison above was the pre-registered one, and it is also open to an objection it cannot answer: *"invalid input must raise"* does not say whether a length mismatch is invalid, while the pin does. The prose and the pin do not carry the same information, so a difference between them need not be a difference of *form*. As a **post-hoc analysis** — added after the pre-registered comparison and reported separately from it — we added a third condition: a prose rule whose scope is explicit, taken verbatim from the benchmark's own description of the chosen convention (for the task above: *"if the lengths differ, raise `ValueError`"*). It was scored on the same population as the second held-out construction — nine tasks on one model, ten on the other — on both the pinned and the held-out input. The general rule and the pin are restricted to that same population below, so that all three columns describe the same tasks:

```
same 9 + 10 tasks       pinned input            held-out input (2nd construction)
                        general  specific  pin   general  specific  pin
claude-sonnet-5          8/9      9/9    9/9      8/9      9/9     9/9
openai:gpt-5.5           9/10    10/10  10/10    10/10    10/10   10/10
```

The scope-explicit rule was not evaluated on the first held-out construction, where only the general rule and the pin were compared (8/9 vs 9/9 and 10/10 vs 10/10 on the same nineteen tasks).

The scope-explicit prose rule matches the pin in every cell, including the one cell where the general rule failed. What this directly shows is that **the improvement we observed over the general rule can also be produced by stating the rule's scope in prose**; it does not show that nothing else about the pin contributes, since no cell remained in which the two could differ.

We therefore state the policy half narrowly, in three parts. A pin is honoured at the input it names (21 of 21 tasks, 210 generations). On a held-out input it was never worse than the general prose rule in any of 38 cells, nor than the scope-explicit rule in the 19 cells where that was measured; we do not generalise beyond those cells. And in those 19 cells its form conferred no measurable advantage over prose stating the same scope. The contribution is the dual use of the decision record: it guides generation, as these results show, and the same record also serves directly as an executable check, whose properties Section 5.2 measures.

---

## 6. Failure Modes

The compliance rates of Sections 5.3 and 5.4 leave a small number of cases unaccounted for. They are not scattered: in every generation we constrained by a prose rule, on both models, the deviations occur through one mechanism. (The disagreements of Section 5.2 are a different matter — misclassification by a single input, not non-compliance — and are not included here.)

### 6.1 The model decides the scope of the rule

Every genuine deviation we observed under a prose rule involved **a rule that does not say which inputs it applies to**, and in each case the model settled that question differently from the rule's author. The rule in question was always the one reading *"invalid input — negative values, out-of-range values, type violations, anything not satisfying the contract — must raise an exception."*

In RQ3 the effect appears as under-application:

```
is_samepatterns(['red','green'], ['a','b','c'])
    the model decides a length mismatch is not invalid input, and returns False
is_woodall(0)
    the model decides 0 is not invalid input, and returns False
```

Both were three-to-two splits across repeated generations, not refusals: the model is not rejecting the rule, it is unstable about whether the rule applies.

The more consequential form is over-application, where the rule's self-assigned scope swallows the territory of another rule. Two recorded decisions in our store are damaged this way, both of them decisions about preserving a numeric type, and we characterise one completely because it reproduces on both models.

**Case: `octagonal_number(3.5)`.** The recorded decision is that the input's numeric type is preserved — the pin expects `29.75`. The store also contains the invalid-input rule. With the type-preservation rule alone, both models produce:

```
def octagonal_number(n):
    return n * (3 * n - 2)
```

With all twelve rules present, both models instead produce:

```
def octagonal_number(n):
    if not isinstance(n, int) or isinstance(n, bool):
        raise TypeError("n must be an int")
    if n < 1:
        raise ValueError("n must be a positive integer")
    return n * (3 * n - 2)
```

The model has classified a float as a *type violation*, which the invalid-input rule then requires it to reject — and in doing so it has contradicted the recorded decision, which exists precisely to say that a float is a legitimate input whose type must be preserved.

A leave-one-out sweep attributes this uniquely. Removing the invalid-input rule from the store restores `29.75`; removing any one of the other ten rules leaves the behaviour broken. The result is identical on both models:

```
                           claude-sonnet-5   openai:gpt-5.5
  type rule alone                29.75            29.75
  all twelve rules             TypeError        TypeError
  all twelve minus invalid-input  29.75            29.75
  all twelve minus any other   TypeError        TypeError   (10 of 10)
```

The second damaged decision behaves the same way but less cleanly, and the difference is instructive. At three generations per condition, two removals appeared to restore it. At nine, only the invalid-input rule does so consistently (9 of 9); the second candidate restores it in 4 of 9, and an unrelated control rule removed as a check leaves it broken in 9 of 9. A sweep of twelve conditions at three samples each will produce a spurious attribution often enough to matter, and the influence of a rule on a decision is not always all-or-nothing.

Neither rule is wrong, and neither mentions the other. The conflict is created by the model's judgement about what *"type violation"* denotes, which is exercised at generation time and recorded nowhere.

Two clarifications about what these runs are. The store presented to the generator consisted of the twelve rules **as prose**; the pin's recorded value `29.75` served only as the detector. What was overturned was therefore a decision expressed to the model as a prose rule, and what caught it was re-running the pin. Whether the same decision expressed to the model *as a pin* would have survived the store is a different question, and **not one this paper answers for these two tasks**: on both, the model's unconstrained default already agrees with the recorded decision, so they fall outside the population of Section 5.4, whose 21 tasks are those where the default disagrees. The survival of pins under the store reported there is a result on a different set of tasks.

We offer this as a candidate mechanism behind the adherence rates reported by CIFE, which measures constraint following at scale but does not analyse why following fails. The prediction it implies is testable: adherence should be markedly worse for constraints whose scope is given by an open predicate than for constraints whose scope is enumerated.

### 6.2 Reconciling rules by their text misses this

A natural way to prevent the case above is to reconcile rules against each other when one is added, which is how TRACE's rule lifecycle works. We asked both models to do exactly that: given the store and a rule being added, name the existing rules that conflict with it.

Presented with the type-preservation rule as the addition, **neither model named the invalid-input rule, in five attempts out of five.**

```
claude-sonnet-5    0/5    (named an unrelated rule once)
openai:gpt-5.5     0/5    (named nothing)
```

This is one conflict, and we do not generalise from it to a rate. What it shows concretely is narrower than a claim about text-level reconciliation as such: with this store, this prompt and five attempts, neither model inferred the conflict from the rule texts, while re-running the recorded input observed the damage directly (`29.75` becomes a `TypeError`). The two rules share no vocabulary, which makes the inference harder, but a reconciler that reasoned about what *type violation* denotes could in principle find it; these did not.

We also measured textual reconciliation more broadly. We wrote one deliberately opposing rule for each of the twelve rules in the store and asked each model, for each pair, which existing rules the addition conflicts with; eleven of the pairs concern rules that govern tasks in our set. Reliability differed sharply by model — nine of eleven pairs identified by one, five of eleven by the other, under an identical prompt and store. A mechanism whose accuracy varies by a factor approaching two across vendors is a fragile place to put a safety property.

These opposing rules are our own constructions and the exercise is artificial; what it establishes is only that text-level reconciliation is not a stable instrument, not how often real rule sets contain conflicts it would miss.

### 6.3 Conflict damage barely occurs at this scale

The rule pair in 6.1 is the only one in our store that damaged a recorded decision — one causal rule, one pair, two tasks — and we did not find it by looking for it. We tried three designs to produce such damage deliberately:

```
store of 11 unrelated rules, on decisions the model would have made anyway
    3 of 51 task-model pairs damaged

rules written specifically to oppose each policy, same population
    3 of 51 pairs damaged (10 of the 12 opposing rules had no effect at all)

same, on decisions that override the model's default
    0 of 19 pairs damaged
```

We had intended to evaluate a sixth step — re-running every stored pin on each append, to catch conflicts as behavioural regressions — and pre-registered a comparison against textual reconciliation. With this few instances of damage the comparison has no power, and we report it as unanswered rather than choosing among the readings it permits.

Part of the explanation is a feature of the evaluation design. Each task carries one decision point and shares no code with any other, so the benchmark excludes interference through shared code and therefore provides limited coverage of the conflicts that may arise as decisions accumulate. It does not exclude interference through the model's reading of the rules — the pair in 6.1 arises that way, between independent functions — and how often that kind occurs at scale is not something this benchmark can say. Evaluating conflict regression properly requires a setting in which many decisions accumulate against a shared codebase, which is also the setting required to test the amortisation claim we have excluded from this paper: two different lines of argument arrive at the same missing experiment.

We do not draw the practical conclusion that conflict regression is unnecessary; a step we could not evaluate is not thereby shown to be idle. What the negative result supports is only this: in this setting, recorded decisions survived an eleven-rule store and most deliberately opposing rules, so too few violations arose to measure what the step would add.

---

## 7. Threats to Validity

### The benchmark

**The tasks are small.** Forty functions derived from MBPP. A decision point in a forty-line function is not a decision point in a system, and nothing here speaks to the latter.

**The decision points were drafted by a model.** They are therefore drawn from the set of points a model can recognise as underdetermined, which may be a proper subset of the points that matter. The mechanical checks constrain the form of a task, not its representativeness.

**Task selection involved judgement that the checks do not capture.** Acceptance rests on the mechanical checks, but a review pass removed a further 21 candidates. Eleven of those can be re-derived from the requirement text or by the current automatic check; the other ten rest on whether a stated convention is one that is actually practised, and an author confirmed each of the ten individually. What remains is that the first pass over all 64 candidates was made by a model: a candidate it wrongly *kept* would not have been surfaced for that review. We have no estimate of that rate.

### Adjustments made after seeing results

We record them in full because each was prompted by an observation rather than planned.

**Repairing reference implementations.** Eight tasks produced an implementation matching neither reference when generated with no policy at all; five were repaired and three removed (Section 4.3). The criterion makes no reference to any method's score, which is what makes it defensible, but the tasks were selected for examination by a measurement.

**Normalising exception types.** Our judgement originally compared raw outputs, so an implementation that raised `ValueError` where the reference raised `TypeError` was scored as having resolved the decision differently — although for every decision point in the benchmark that concerns exceptions, the decision is *whether* to raise, not which exception to raise. We introduced normalisation and confirmed that no task in the benchmark has exception type as its decision point. Re-judgement used the stored generations; no new sampling was performed. The change moved one baseline classification and corrected four apparent policy violations that were in fact compliant.

**A known reference defect left unrepaired.** The check of Section 5.2 exposed a defect in one task's reference implementation (`mbpp/36`, digit extraction for `p > q`), which the repair criterion of Section 4.3 cannot detect. We chose not to repair it: doing so would be another post-hoc adjustment made after seeing a result, and the figure it affects — ten of fourteen disagreements on one model — is more useful reported as measured. It is recorded as an erratum with the benchmark and will be fixed in the next version. Its reach across the results is as follows. RQ1 counts the task as reached on both models; not counting the defect-derived success as a success gives 36 of 40 on the model whose pin identifies the defect (90.0 percent) and leaves the other unchanged. RQ2's disagreement rate on that model falls from 14 of 350 to 4 of 340 (1.2 percent); the other model's is unchanged, its pin for the task using an input on which the defect does not arise. RQ3 and RQ4 are unaffected: the task's decision coincides with both models' default and so lies outside their populations. Section 6 does not use it.

**Replacing the false-alarm measure.** An earlier draft reported, as RQ2's false-alarm count, that the pin's input exposed no difference between the two *references* on the benchmark's distractor inputs. That is a re-run of check

E3 and says nothing about implementations the pin might wrongly reject; we withdrew it and replaced it with the agreement measurement of Section 5.2.

**Reselecting the held-out inputs.** Our first construction of the held-out input of Section 5.4 required only that it discriminate the decision. Seeing that the prose rule then passed in all but one cell, we added the requirement that the model's unconstrained output on the held-out input land on the opposite convention, and reselected from a pool of 296 candidates. The outcome was unchanged, and we report both runs. The redesign was prompted by a result, but it tightens the condition symmetrically: the new requirement makes the input harder for the pin and for the prose rule alike.

**The first two adjustments raise the reported rates.** The figures before and after are:

```
                          43 tasks, raw   40 tasks, raw   40 tasks, normalised
RQ1 (1)+(1')  sonnet       90.7%           92.5%           92.5%
              gpt          90.7%           92.5%           92.5%
RQ3           sonnet       94% (15/16)*    70.0% (7/10)    90.0% (9/10)
              gpt          94% (15/16)*    75.0% (9/12)    90.9% (10/11)**
```

* after reading the generated code by hand, which is what the normalisation later made mechanical; ** one task left the RQ3 population when its convention text was corrected (Section 4.3). The conclusions of Sections 5.1, 5.3 and 5.4 hold under every column; the full trail is in the accompanying record.

### The measurements

**Two models.** Results may depend on this pair. The per-model differences in Section 5.1 — which stage of pin construction fails — are the observations most exposed to this.

**Step (1') does not certify what it appears to.** Its self-validation confirms that a proposed input separates the model's own two candidate implementations. It does not confirm that the input separates the intended decision, and we observed a case where it does not (Section 5.2). Independently, the precision of step (1') is low: among the points it certifies, 37 to 40 percent concern the injected decision.

This limitation is a known one in a neighbouring setting, and we arrived at it independently. Richter and Papadakis observe that clustering-based clarification asks about whatever the model's own samples happen to differ on, so that two wrong implementations differing only in edge-case handling produce a question about edge cases rather than about the real divergence; on their MBPP variants a clarification turn raises detrimental collapse to between 37 and 77 percent. Our step (1') is not clustering — the point it pins comes from step (1)'s enumeration, not from observed variance — but the input that pins it is still constructed from a pair of implementations the model wrote, and inherits the same exposure. The `mbpp/62` case in Section 5.2 is an instance. Deriving the input by symbolic execution, as ARHF does, would make the input independent of the model's sampling, but not the candidates' fidelity to the intended reading; that remains the human check at step (3).

**Small denominators in the adverse conditions.** RQ3 and RQ4 rest on 10 and 11 tasks, because these are the tasks on which a recorded decision can be distinguished from the model's default at all. The perfect scores in Section 5.4 are 21 of 21 tasks, not rates estimated from many.

**A generation failure mode contaminated an early count.** In one condition, generated code occasionally failed to load, and our classifier mapped that to the same value as a changed decision. Two of four apparent conflict damages were of this kind (Section 6.3 reports corrected counts). The rate of such failures was 4 of 81 in the affected condition and zero in every other condition and model.

### Scope of the conclusions

**Representation dependence, and this is the serious one.** Our design assumes, following Richter and Papadakis, that resampling a requirement does not by itself expose an underdetermined point, which is why a decision point must be injected and reference implementations supplied. VeriMed applies the same resampling procedure but compares *logical encodings* rather than code, and finds 12 of 64 requirements producing more than one encoding, with 44.0 percent pairwise agreement across sampled formalisations. If disagreement is visible in one representation and not another, then the premise of our design — and by extension the necessity of step (1') — may hold only for code.

**Three claims about the pin as policy, and the scope of each.** (i) *Compliance at the pinned input* — 21 of 21 tasks; the graded input is the shown input, so this measures compliance with an instruction, not generalisation. (ii) *Generalisation to an unnamed input* — no cell of failure in the 38 measured; nothing is claimed beyond those cells. (iii) *Superiority over a prose rule* — not established: the improvement over the general rule was reproduced by stating the rule's scope in prose (Section 5.4), leaving no cell in which the pin's form could show an effect.

**One rule pair is one rule pair.** Section 6.1 characterises a single interaction — one causal rule, one pair, two damaged tasks — completely and across both models. It is not a rate, and the observation in Section 6.2 that text-level reconciliation did not find it rests on that single pair together with a pre-registered comparison we report as underpowered.

**The distance to the motivating goal.** The motivation of Section 1 is a development process in which human review is spent only on questions not yet decided, because decided ones are held by executable records. This paper evaluates the record, not the process, and the parts of the process it leaves unevaluated are the parts that would make the record pay for itself. They are scattered through the paper and we gather them here. Recognising that a new requirement raises a point already in the store is not evaluated; the relevant pin was supplied directly (Section 3.2). Carrying a pin over when the function or its arguments differ from the recorded case is not addressed. Updating pins when a decision changes is not addressed. The human cost of step (3), and of judging whether a constructed candidate pair is faithful to the point named, is not measured (Section 3.3). And whether a store of pins reduces human cost over time — the amortisation the motivation assumes — is not evaluated; it requires decisions accumulating against shared code, the same setting missing from Section 6.3. These are not caveats on the results above but the research that remains between them and the goal.

---

## 8. Conclusion

Underspecification in natural-language requirements is not a defect to be eliminated; models resolve it silently because they have no other option, and the resolution goes unrecorded. We inserted two steps into the ordinary practice of writing acceptance examples before implementing — enumerating the underspecified points, and verifying each candidate input against a pair of implementations constructed to differ in that point alone — and recorded what they produce as a **decision pin**: the named point, the verified input, and the value the person chose. It is a single artifact that constrains later generation and decides compliance, written before the target implementation is generated rather than after a failure.

Across 40 tasks and two models, an input separating the two reference readings was obtained for 37 of 40 decision points on each model, of which 35 and 36 survive the comparison rule and 34 and 36 identify the intended decision rather than an unrelated difference. Used as checks on 703 independently generated implementations, pins agreed with the benchmark's own single-input classification on 96 to 97 percent; among the 24 disagreements the benchmark's input was misled by a coincidence on fourteen and the pin on fourteen, four of them the same implementations — the pin's errors being ten from a defective reference value and four from a crash the comparison rule cannot tell from a raise — which is to say that single-input classification is fragile in both directions. Used as constraints on generation, a recorded decision was honoured at its pinned input in all 210 generations, with and without eleven unrelated rules present. On held-out inputs it was never worse than a general prose rule in 38 cells, and no better than a prose rule stating the same scope in the 19 cells where we compared them. The results support using the same decision record for both generation guidance and executable checking, without establishing an advantage over scope-explicit prose in generation adherence.

The failures under prose rules have a single shape. Every deviation we observed in generation constrained by a prose rule traced to the same thing: a rule that leaves its own scope to the reader, and a model that sets that scope for itself. In one case the effect is severe enough to silently overturn a different recorded decision, and the two rules involved share no vocabulary; asked to reconcile them by their text, neither model found it, while re-running the recorded input did. We could not, however, produce enough instances of such damage in this setting to evaluate a regression step built on that observation. Our tasks share no code, so the benchmark gives limited coverage of the conflicts that accumulating decisions may produce — the same limitation that keeps us from speaking to whether recording decisions repays its cost over time. Both point at the same experiment we have not run.

---

---

## Data and Artifact Availability

The benchmark (40 tasks with paired reference implementations, probes and distractors, together with the audit of rejected candidates and the recorded erratum), all experiment runners, the raw generation logs, the pre-registered designs for each experiment, and the record of every adjustment made after seeing a result will be made publicly available with the final version of this paper. The number audit that cross-checks every figure in this paper against those logs is included and can be re-run.

## Use of Generative AI

Following arXiv's policy that significant use of text-to-text generative AI be reported as part of the methodology, we state the division of labour precisely. This paper was produced with substantial use of a large language model (Claude, Anthropic).

*Performed by the author:* the research question and the motivating goal; the design of each experiment and the pre-registration of its judgement rules before execution; the decision to withdraw claims whose pre-registered criteria were not met; the confirmation of each of the ten benchmark candidates rejected on judgement; the choice not to repair a discovered reference defect and to report it instead; the selection of related work to read in full; and the acceptance or rejection of every change proposed in ten rounds of review.

*Performed by the model under the author's direction:* implementation of the experiment harness and analysis scripts; execution of all runs; extraction and reading of the full text of cited papers and the drafting of the per-paper notes on which Section 2 is based; drafting of the manuscript in English; and the number audit.

*Language:* the author is not a native speaker of English. The English of this manuscript was drafted and edited by the model from the author's decisions and notes in Japanese.

All measurements, the trail of corrections, and the audit are released so that the division above can be checked against the artefacts rather than taken on trust.